\documentclass[journal=ancac3,manuscript=letter,layout=twocolumn]{achemso} 
\setkeys{acs}{articletitle = true}
\usepackage[version=3]{mhchem}
\usepackage[T1]{fontenc}
\usepackage{tabu}
\usepackage{amsmath}
\usepackage{amsfonts}
\usepackage{mathtools}
\usepackage{bm}
\usepackage{esvect}
\usepackage{fixltx2e}
\usepackage{hyperref}
\usepackage{amsfonts,amssymb,amstext,amsmath,amsthm,verbatim,times,cancel,epsfig}
\usepackage{tikz,pgfplots}
\usepackage{caption}
\usepackage[english]{babel}
\usepackage{natbib}
\setcitestyle{square}
\usepackage{floatflt,epsfig}
\usepackage{xcolor}

\author{Giuseppe Emanuele Lio}
\affiliation[University of Calabria]
{University of Calabria, Physics Department, \\87036 Arcavacata di Rende (CS), Italy}
\alsoaffiliation{CNR-Nanotec, Cosenza 87036 Arcavacata di Rende (CS), Italy }
\author{Antonio Ferraro}
\affiliation[University of Calabria]
{University of Calabria, Physics Department, \\87036 Arcavacata di Rende (CS), Italy}
\alsoaffiliation{CNR-Nanotec, Cosenza 87036 Arcavacata di Rende (CS), Italy }
\email{antonio.ferraro@unical.it}
\author{Michele Giocondo}
\affiliation[CNR]
{CNR-Nanotec, Cosenza 87036 Arcavacata di Rende (CS), Italy }
\author{Roberto Caputo}
\affiliation[University of Calabria]
{University of Calabria, Physics Department, \\87036 Arcavacata di Rende (CS), Italy}
\alsoaffiliation{CNR-Nanotec, Cosenza 87036 Arcavacata di Rende (CS), Italy }
\alsoaffiliation{Institute of Fundamental and Frontier Sciences, University of Electronic Science and Technology of China, Chengdu 610054}
\email{roberto.caputo@unical.it}
\author{Antonio De Luca}
\affiliation[University of Calabria]
{University of Calabria, Physics Department, \\87036 Arcavacata di Rende (CS), Italy}
\alsoaffiliation{CNR-Nanotec, Cosenza 87036 Arcavacata di Rende (CS), Italy }
\title[An \textsf{achemso} demo]
     {Color Tuning by Spontaneous Propagation of Gap Surface Plasmons in Epsilon Near Zero Nano-Cavity}
\keywords{ENZ, Ellipsometry, Optical nano-cavity, Plasmonics, sensing}

\begin{document}
\maketitle

\begin{abstract}
This work presents numerical and experimental results of plasmonic Metal-Insulator optical nano-cavities. The systems are composed of Silver as metal, polyvinylpyrrolidone or indium tin oxide or zinc oxide as insulator. The proposed nano-cavities exhibit extraordinary effects as colors changing in function of the incident/view angles, enhancement of the experimental real and imaginary parts of the pseudo dielectric constant, an extraordinary transmission of 50$\%$ and zero reflection for both polarizations at the resonant wavelengths. These results lead to the formation and propagation of surface plasmon polaritons and their hybridization in gap surface plasmons. The concurrence of these effects allow studying the Goos-H\"anchen shift by exploiting the very narrow $\Delta$ and $\Psi$ ellipsometer parameters. Thank to their optical properties, the proposed nano-cavities could be applied in several fields such tunable color filters, optics, photonics and sensing.
\end{abstract}

\section{Introduction}
Nanotechnologies based on plasmonic and photonics effects, thanks to the interaction between light and matter at the nanoscale, are involved in several applied field as enhanced Raman spectroscopy, bio-sensing, transistor  \cite{xu1999spectroscopy, chang2007single, ward2007electromigrated, li2011three, de2012tailoring, berweger2012light}. Furthermore, the light propagation into nano-guided system could be exploited for the fabrication of subwavelength optics and nano-optical devices \cite{barnes2003surface, de2014tuneable, lio2019integration}. A Polariton is an electromagnetic wave coupled to a polarization excitation in matter \cite{ruppin1970optical, mills1974polaritons, halevi1978polariton}. When this coupled excitation occurs at the interface between two media, it is called a surface polariton\cite{zayats2005nano}. The latter is an evanescent electromagnetic wave that propagates along the interface with amplitude decaying exponentially into the two media. When one of the media is a metal and the other a dielectric, the propagating waves are referred as surface plasmon polaritons (SPPs) . Since the discovery of the SPPs, the plasmon resonance at small gaps in metallic subwavelength structures has attracted interest in terms of spectroscopy behavior.
An open research field, in particular, is represented by systems composed of Metal/Insulator/Metal (MIM) that are of scientific interest because of the dielectric singularities that appear in the anisotropic permittivity due to $\varepsilon_{NZ}$ (NZ stands for near zero) conditions\cite{maas2013experimental, caligiuri2016metal, caligiuri2016dielectric}. When the dielectric layer is thicker (hundreds of nanometers) than the metal layers (tens of nanometers), the system behaves as a nano-optical cavity. These systems allow efficient concentration and confinement of electromagnetic energy and can be used for numerous applications as light emission and detection \cite{vahala2003optical},  photovoltaic \cite{grover2011applicability, guo2014metallic, grover2012engineering}, Bragg reflector \cite{hosseini2006low, zhang2009subwavelength}, plasmonic devices \cite{han2007surface, wang2011tunable}, for optics and the related applications \cite{hill2009lasing, zhang2009subwavelength, choo2012nanofocusing}, telecommunications \cite{wang2011tunable, kim2002v}, and others \cite{thissen2010optical, banerjee2009nanotubular, caligiuri2018planar}.
    Under a physical point of view, when metal/insulator and insulator/metal interfaces are close to each other, the dispersion curve of the single interface splits into high and low energy modes. \cite{miyazaki2006squeezing}  This gives the possibility to excite the surface plasmon by a free space wave without momentum matching simply by perpendicular incidence to the bare metallic end-face of the MIM \cite{stegeman1983excitation}. At the interfaces of the MIM, spontaneous surface plasmon polaritons (SPPs) arise whose electric field overlaps within the insulator layer.  However, in the MIM system due to the matching  of the field symmetry, only the low energy mode is excited (using a p-polarized electromagnetic wave). In case of a MIMI system, it is instead possible to excite both low and high (guided) energy modes, respectively with p- and s-polarized light. \cite{zia2004geometries, dereshgi2016large, sharma2017surface}. 
In presence of the hybridization of the SPPs excited at the two metal interface and decayed in the thick dielectric layer, it has been numerically and experimentally demonstrated the formation of gap surface plasmons (GSPs) that are related to high energy modes\cite{miyazaki2006squeezing, miyazaki2006controlled, kuttge2009dispersion, neutens2009electrical, todisco2016toward}. 
Although several studies of GSPs are reported in literature, the problem of reproducible fabrication of small nanometric gaps persists. In this manuscript, an easy way is proposed to design and fabricate large-area plasmonic cavity resonator in which a dielectric layer of a few hundred nanometers is employed as a gap between the metal layers. The proposed nano-cavity has been designed using numerical ellipsometer analysis (NEA), which is able to fit the ellipsometer spectroscopic measurements retrieving optical constant of materials\cite{Lio2019cavities}. In order to fabricate the real nano-cavities, silver (Ag) is exploited as the metal and, for the insulator, three different materials are considered (polyvinylpyrrolidone (PVP), indium tin oxide (ITO) and zinc oxide (ZnO)). 
The ellipsometric characterization of the proposed systems reveals very interesting optical features such as extraordinary transmittance\cite{han2011plasmon} and $0\%$ reflectance for both exciting polarizations (p and s). Furthermore, the ellipsometric parameters $\Psi$ and $\Delta$ are measured. These are related to the reflectance ratio $\rho=tan(\Psi)e^{-i\Delta}$, that includes the amplitude component, calculated as $\Psi= arctan(r_p/r_s)$ with $r_{p(s)}$ being the $\textit{Re}\left \{\sqrt{R_{p(s)}}\right \}$, and the phase difference evaluated as $\Delta=\phi_p - \phi_s$\cite{Lio2019cavities}. In presence of the plasmonic resonance, due to the SPPs and GSPs formation, these parameters show a particular sigmoidal behavior leading to the enhancement of the real and imaginary part of the pseudo dielectric constant. The constant is calculated as $<\tilde{\varepsilon}>=<\varepsilon_1>-i<\varepsilon_2> = sin^{2} \theta_{inc} [1-tan^{2} \theta_{inc} (\frac{1-\rho}{1+\rho})^{2}]$\cite{azzam1978ellipsometry}, where $\theta_{inc}$ is the incident angle. Due to the presence of $\rho$ in the previous expression, $<\tilde{\varepsilon}>$  is directly related to $\Psi$ and $\Delta$. Under specific conditions, $<\varepsilon_2>$ can reveal a gain behavior of the system reaching very high absolute values  (500) for the MIMI system with ZnO as insulator. Due to their plasmonic behavior, the fabricated nano-cavities present a macroscopic optical effect characterized by different colors when incident light and viewing angles are tuned\cite{xu2011structural, chen2012plasmonic, gu2015color, segal2016hybridization, caligiuri2020angle}. 
Finally, behavior of $\Delta$ and $\Psi$ evidence the presence of a spontaneous Goos-H\"anchen shift effect\cite{snyder1976goos, wild1982goos, lai1986goos}. This effect is observed as a small lateral shift between the p and s directions of linearly polarized light undergoing total internal reflection. The counterpart of this effect for circular and elliptical polarization is the Imbert-Fedorov effect\cite{de2001evanescent}. The Goos-H\"anchen effect arises when both waves are reflected from the surface and undergo different phase shifts, that lead to the lateral shift of the probe beam. As demonstrated in previous works\cite{foster2004methods, foster2007goos}, the phase shift appears when resonant modes are induced within an optical dielectric cavity, as in the proposed MIMI.  In fact, in presence of a multi-layered systems ending with a dielectric layer, as in case of a Bragg reflector or a dielectric mirror, it is possible to evaluate the Goos-H\"anchen shift at the resonant wavelength as $\Delta_{s/p} (\theta_{inc}) = 2 t_{1}tan(\theta_{inc})$. In the previous, $t_1$ is the thickness of first dielectric layer in the MIMI configuration. The proposed systems, exhibiting this effect, can be involved in a wide scenario of scientific research such as nanophotonics applications\cite{wild1982goos}, sensitive detection of biological molecules \cite{jiang2017multifunctional, sreekanth2013sensitivity, sreekanth2019phase}, or for the phenomenological study of the lateral reflection for both polarizations\cite{Merano07}.

\section{Results and discussion} 
\subsection{Extraordinary transmission in polymeric MIM}
\begin{figure}[H]
\begin{center}
\includegraphics[width=0.6\columnwidth]{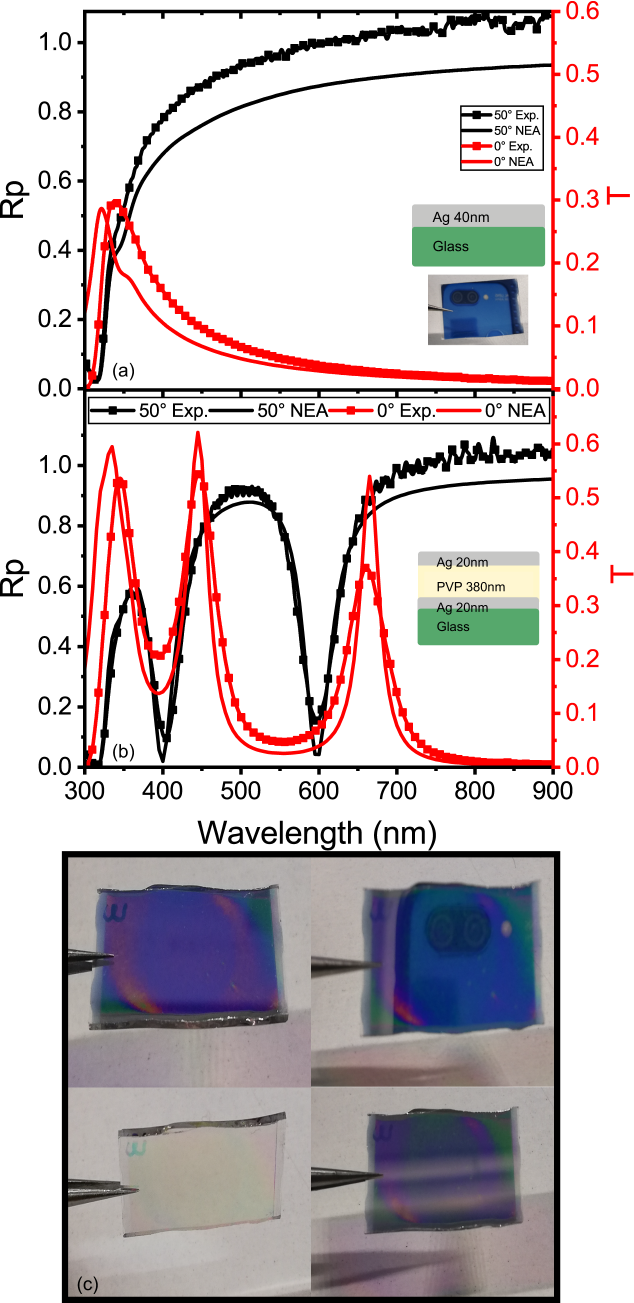}
  \caption{a) Numerical and experimental reflectance (p-pol) at $50^\circ$ and transmittance at $0^\circ$ of a silver with thickness of $40nm$. The insets report a schematic view of the fabricated system with the thickness of each layer and a real picture where the incident light is completely reflected. b) Numerical and experimental reflectance (p-pol) at $50^\circ$ and transmittance at $0^\circ$ of MIM nano cavity composed of Ag/PVP/Ag. The insets report a schematic view of the fabricated system with the thickness of each layer. c) Pictures showing colors dependence on the incident/view angles.}
  \label{1}
  \end{center}
\end{figure}

In order to fully investigate, both numerically and experimentally, the extraordinary features of the fabricated nano-cavities, see methods for information about the fabrication, we first considered a silver layer with thickness of $40nm$ deposited on glass substrate. As expected, the transmittance presents a peak only in the UV range at around $330nm$ with a maximum of the $30\%$, while the other wavelengths (solid lines with symbol) are reflected. The experimental results agree very well with the numerical one (solid lines)\cite{Lio2019cavities} as reported in Fig. \ref{1}a.
Then, the Ag thickness has been divided in two parallel layer of $20nm$ each, placing between them a PVP layer of $380nm$. A MIM nano-cavity Ag/PVP/Ag has been created which presents an interesting optical behavior. In fact, in Fig. \ref{1}b it is possible to recognize a three main peaks/dips in transmittance/reflectance at specific resonant  wavelengths namely in this case at blue, green and red. Moreover, when the MIM is placed at specific small angles it appears transparent -bottom-left of Fig. \ref{1}c- while at higher angles it appears colored both in reflection or transmission, as shown in the other sub-pictures of Fig. \ref{1}c. This angle sensitive could arise from the plasmonic nature at the resonant wavelength due to the multiple SPPs and GSPs formation at the two metal surfaces and inside the PVP layer, respectively, as explained in the previous paragraph. 
The ellipsometer parameters $\Psi$ and $\Delta$, reported in the supplementary figure S1, confirm that the fabricate Ag/PVP/Ag  MIM is, also, extremely sensible to the polarization (p or s-pol) of the incident wave. These parameters show a giant enhanced Goos-H\"anchen shifts \cite{sreekanth2018giant}. Even if the fabrication process of the MIM can lead to some inhomegenities, the agreement between numerical and experimental results is quite satisfactory. 
Using polymer as insulator in MIM systems give the opportunity to create very thick insulator layer (hundreds of nanometers), hence many peaks/dips are present in the visible range according to the relation dispersion of a MIM, as shown in Fig \ref{1}b. However the PVP layer can be damaged during the second Ag layer sputtering deposition. In order to realize systems more stable than the PVP one, we numerically designed and experimentally fabricated nano-cavities using two solid dielectrics: ITO and ZnO.
\subsection{ITO MIMI characterization}
\begin{figure}[H]
\begin{center}
\includegraphics[width=1\columnwidth]{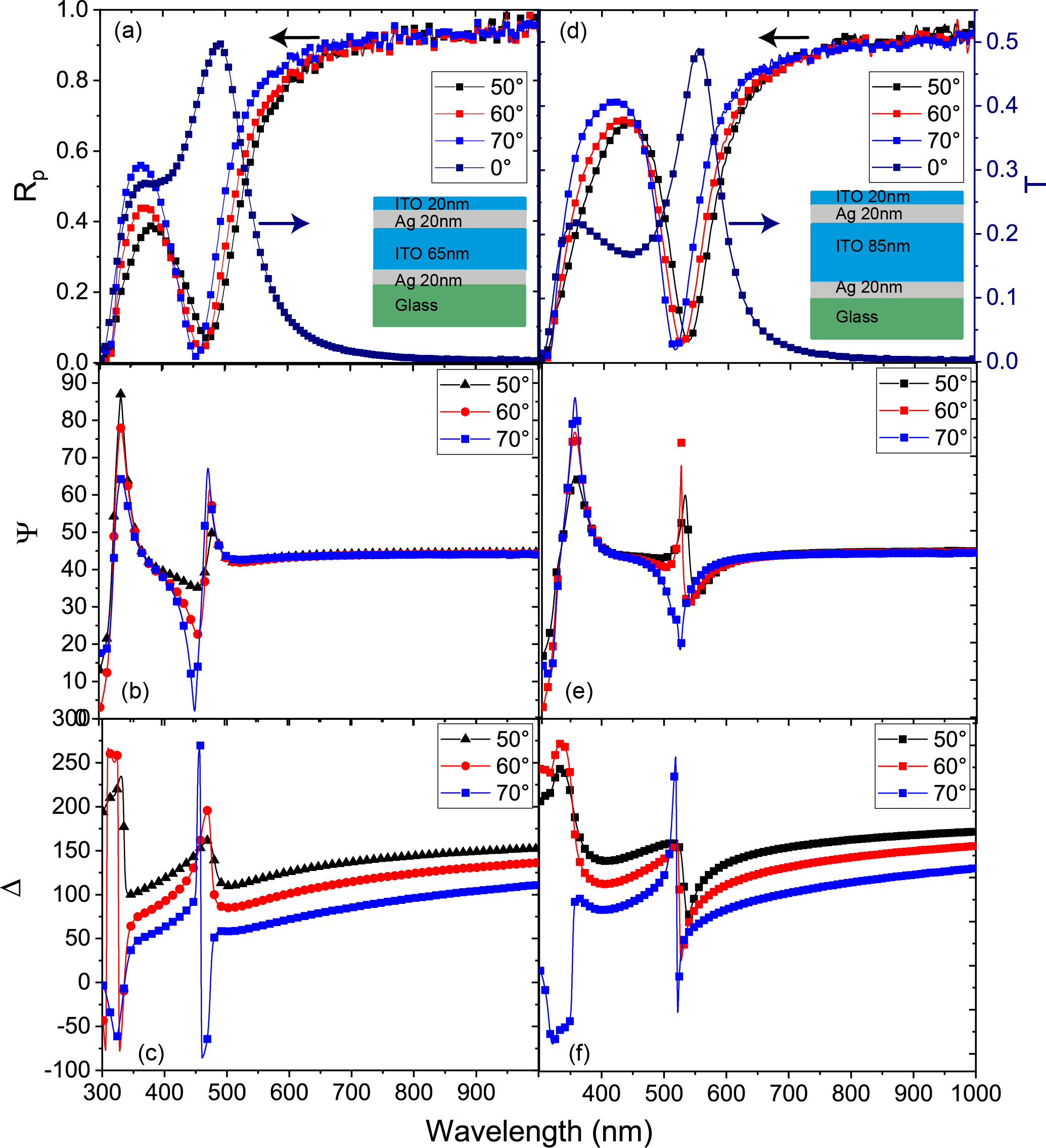}
  \caption {a) Reflectances (p-pol), and transmittance at different incident angles for the MIMI (Ag/ITO/Ag/ITO) made by the ITO layer $t_{1}=65nm$, the inset reports a schematic view of the whole fabricated system with the related thickness of each layer. b,c) $\Psi$ and $\Delta$ for that plasmonic nano-cavity. d) Reflectances (p-pol), and transmittance at different incident angles for the MIMI composed of the ITO layer of $t_{1}=85nm$, the inset shows the whole fabricated system with the new layers thickness. d,e) $\Psi$ and $\Delta$ for the second MIMI.}
  \label{2}
  \end{center}
\end{figure}

\begin{figure}[H]
\begin{center}
\includegraphics[width=1\columnwidth]{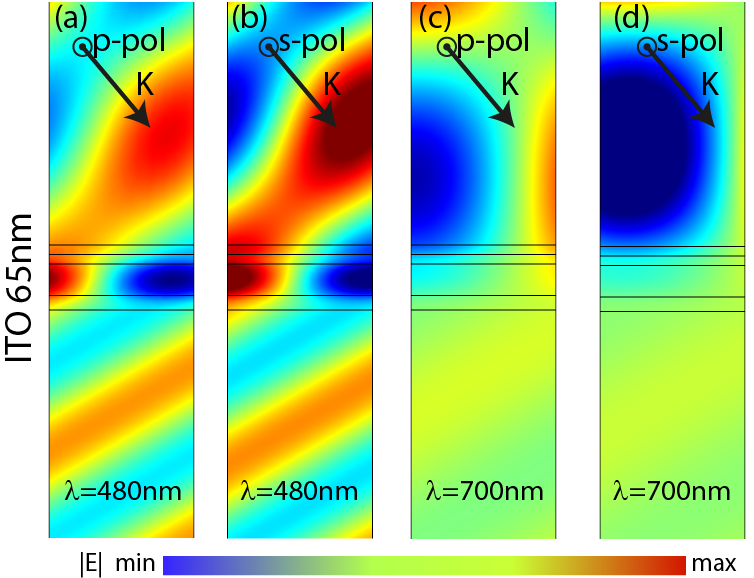}
  \caption{Electric field maps for the MIMI composed of Ag/ITO/Ag/ITO with thicknesses of $20/65/20/20nm$ respectively. a , b) at the resonant wavelength of  $\lambda=480nm$ for p and s-polarization where the resonant mode inside the ITO nano-cavity is shown. c, d) Out of resonance at $\lambda=700nm$ for both polarizations.}
  \label{3bis}
  \end{center}
\end{figure}

\begin{figure}[h]
\begin{center}
\includegraphics[width=1\columnwidth]{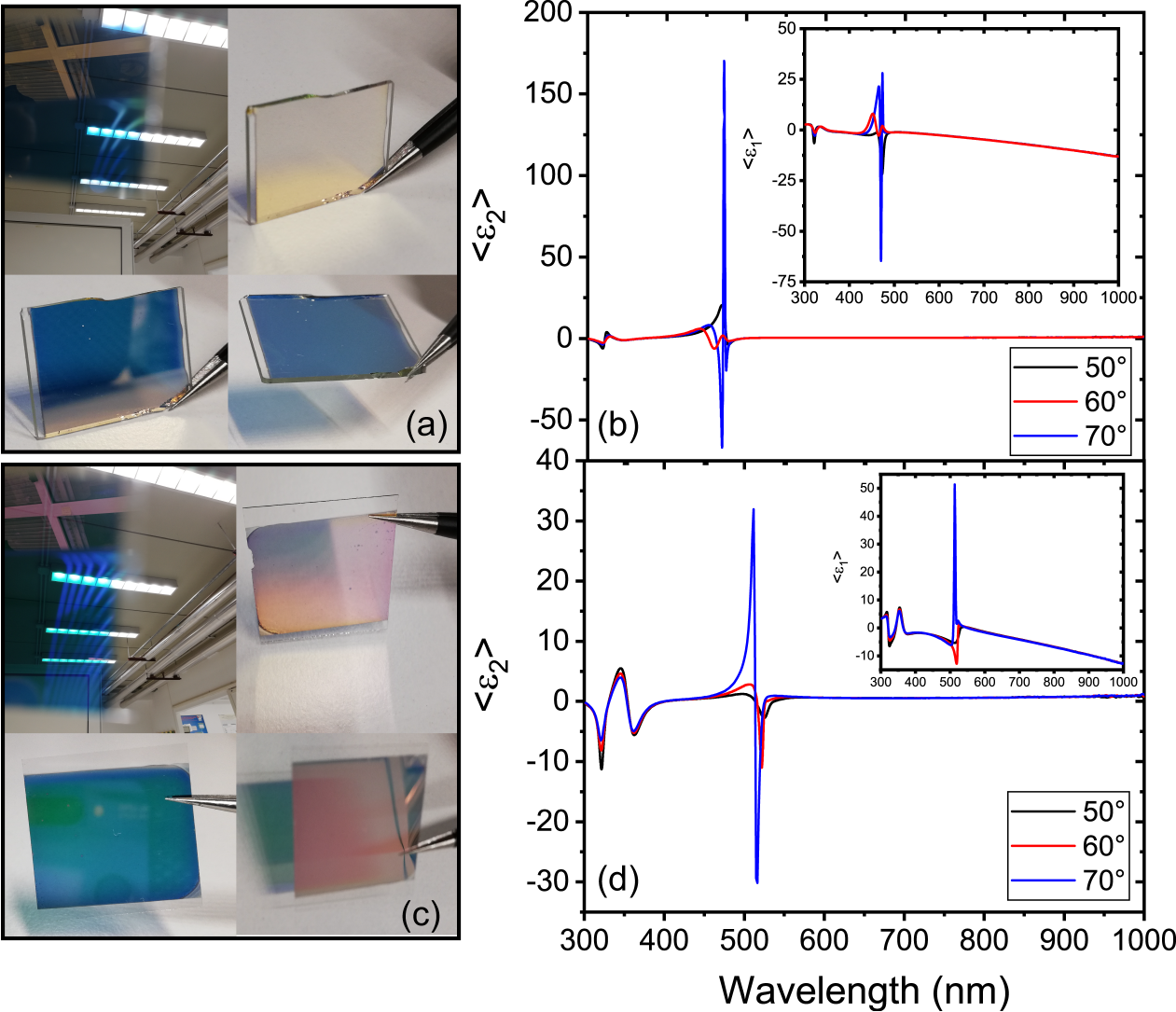}
  \caption{a) The pictures are referred to the MIMI nano-cavity with the ITO thickness equal to $65nm$. The color of the MIMI passes from blue to gold when observed in transmission or in reflection, respectively. b) The graph reports the $<\varepsilon_2>$ at three different incident angles while the inset depicts the $<\varepsilon_1>$.c) The pictures are referred to the second fabricated MIMI  (ITO $t_{1}=85nm$). The nano-cavity color passes from green to purple when it is observed in transmission or in reflection, respectively. d) The graph shows the $<\varepsilon_2>$ at three different incident angles, while the inset shows the $<\varepsilon_1>$ behavior.}
  \label{3}
  \end{center}
\end{figure}

During the numerical designs and the experimental investigation, it has been observed a substantial difference between a classical MIM and a new MIMI metamaterial configuration. The latter presents an accentuate sigmoidal shape in $\Psi$ and $\Delta$ enabling the related study of the plasmonic excitation effects and the Goos-H\"anchen shift, while the MIM does not present this feature.  The comparison between  numerical reflectances for both polarizations, $\Psi$, and $\Delta$ parameters for a MIM and MIMI is shown in Fig. S2. 
The first realized Ag/ITO/Ag/ITO MIMI is composed of ITO layers of thickness $t_{1}=65nm$ and $t_{2}=20nm$, respectively (see inset of Fig. \ref{2}a). 
The experimental measurements show transmittance of 50$\%$ at $0^\circ$ at the resonance wavelength of $\lambda=500nm$, see Fig. \ref{2}a. On the other hand, the reflectance (p-polarization) at $50^\circ$, $60^\circ$ and $70^\circ$ is almost zero around the resonance wavelength $\lambda=480nm$. The MIMI reflectance under s-polarization, reported in Fig.S3a, shows two dips close to the same resonant wavelength of p-pol leading to the particular feature in  $\Psi$ and $\Delta$ parameters shown in Fig. \ref{2}b, c. In fact, $\Psi$ and $\Delta$ present a drastic change in their curves with an accentuate sigmoidal shape at $\lambda=480nm$, enhanced for this particular MIMI at $70^\circ$ see Fig. \ref{2}b,c (blue curve). 
Hence, the $\Psi$ and $\Delta$ parameters highlight the presence of plasmonic excitation for high and low energy modes at the resonant wavelength especially if the system is impinged from an electromagnetic wave with an incident angle of $70^\circ$.
For the second realized Ag/ITO/Ag/ITO MIMI, the first ITO layer is increased to a thickness $t_{1}=85nm$ (see inset of Fig. \ref{2}d) while  the thickness of the other layer remains unchanged. 
The increase of the ITO layer of only $20 nm$ leads to the shift of $50 nm$ of the resonant wavelengths, where the main transmitted peak is now at $\lambda=550nm$  with again a value of almost 50$\%$, the reflectance dips (p-polarization) at $50^\circ$, $60^\circ$ and $70^\circ$ is almost zero around the resonance wavelength $\lambda=532nm$, see Fig. \ref{2}d. The reflectance  for  s-pol is reported in Fig S4a. The parameters $\Psi$ and $\Delta$ for this thicker ITO  layer ($t_{1}=85nm$) exhibit the same behavior of the thinner one -see  Figs. \ref{2}e,f- namely for an incident angle of $70^\circ$ $\Delta$ presents a sigmoidal shape with an high amplitude. This feature can be exploited in a future applications related to the Goss-Hanchen shift \cite{sreekanth2013sensitivity, sreekanth2019phase}. Hence, it is possible to tune easily the resonant wavelengths without affecting the performance of the system.
In order to support the experimental results the reflectances (p and s-pol), transmittance,$\Psi$ and $\Delta$ have been numerically evaluated and reported in Fig S2b,c and S3b,c. 
A further investigation has been done using COMSOL Multiphysics, indeed, it has been studied how the electromagnetic field (\textbf{E}) interacts with the entire system producing the coupled plasmons. Fig \ref{3bis} shows the \textbf{E}, at resonance ($\lambda=480nm$) and out of resonance ($\lambda=700nm$), for the MIMI realized with the ITO layer of $65nm$. In particular, it is possible to observe that most part of the \textbf{E} is confined in the nano-cavity dielectric layer exciting a single (high or low energy) mode while the remaining part of \textbf{E} passes through the system, see Fig \ref{3bis}a, b. 
It has been highlighted that for p-pol the main low energy mode is generated from the SPPs hybridization that decays inside the dielectric cavity, while the s-pol wave excites high energy mode (GSP). 
Instead, out of the resonance condition the electromagnetic field is totally reflected with no interaction with the thicker ITO layer, see Fig \ref{3bis}c, d. The electric field maps related to the second fabricated MIMI (ITO $t_{1}=85nm$) are illustrated in the supplementary information in figure S5. 
The described plasmonic coupling between the incident wave and the system is the main feature of the proposed nano-cavities. In order to investigation how this effect is strongly related to different color changing as function of the incident/view angles, experimental measurements of the real and imaginary part of the pseudo dielectric constants has been done. These quantities are identified in the following as $<\varepsilon_1>$ and $<\varepsilon_2>$. 
Fig \ref{3}a,b show the pictures of the two MIMI with the ITO $t_{1}=65nm$ and $t_{1}=85nm$, respectively. The first one shows different colors when it is observed at different angles, how it is shown in bottom and upper-right pictures of Fig. \ref{3}a. Instead, at small angles the system permits to see through it a blue color while the reflected light appears gold, see upper-left picture of Fig. \ref{3}a. Then, we focused our attention on $<\varepsilon_2>$ for understanding the gain and loss optical behavior, related to the GSP and SPP formation, respectively. The experimental results present an high and low energy mode excited at the resonant wavelength ($\lambda = 480nm$) for incident angle of $70^\circ$, while excited modes (the value is less than the first ones) for each measured incident angles is observed at $\lambda= 327nm$ representing the Ferrel-Berreman mode of Silver \cite{ferrellBerreman2014}, see Fig\ref{3}b.
The second MIMI present different colors passing from the green to the purple due to the increased thickness as illustrated in bottom and upper-right pictures of Fig. \ref{3}c. At small angles, now,  the system allows seeing through it with a green color while the reflected light appears purple, see upper-left picture of Fig. \ref{3}c. The experimental $<\varepsilon_2>$, reported in Fig. \ref{3}d, confirm this optical behavior due to the presence again of an high and low energy mode around the resonant wavelength ($\lambda = 532nm$) for incident angle of $70^\circ$, while a moderate second excited mode is observed at $\lambda= 380nm$ for each measured incident angles. Moreover, the Ferrel-Berreman mode is still present. 
Therefore, these measurements could be used to recognize in a fast and effective way the excitation of a GSP as a gain and of the SPP as a loss in $<\varepsilon_2>$. In fact, the $<\varepsilon_2>$ measurements are largely used to explain the presence of gain or loss effects in plasmonic materials \cite{de2013low, khurgin2012reflecting, sukhorukov2014nonlinear, johns2016role,koenderink2019plasmon, song2017probing, caligiuri2017resonant}. In the present case, this has been used to trace the gain/loss condition of the MIMI (related to the high/low modes) in presence of a s- or p-polarized excitation respetively.
\subsection{ZnO MIMI characterization}
\begin{figure}[H]
\begin{center}
\includegraphics[width=0.8\columnwidth]{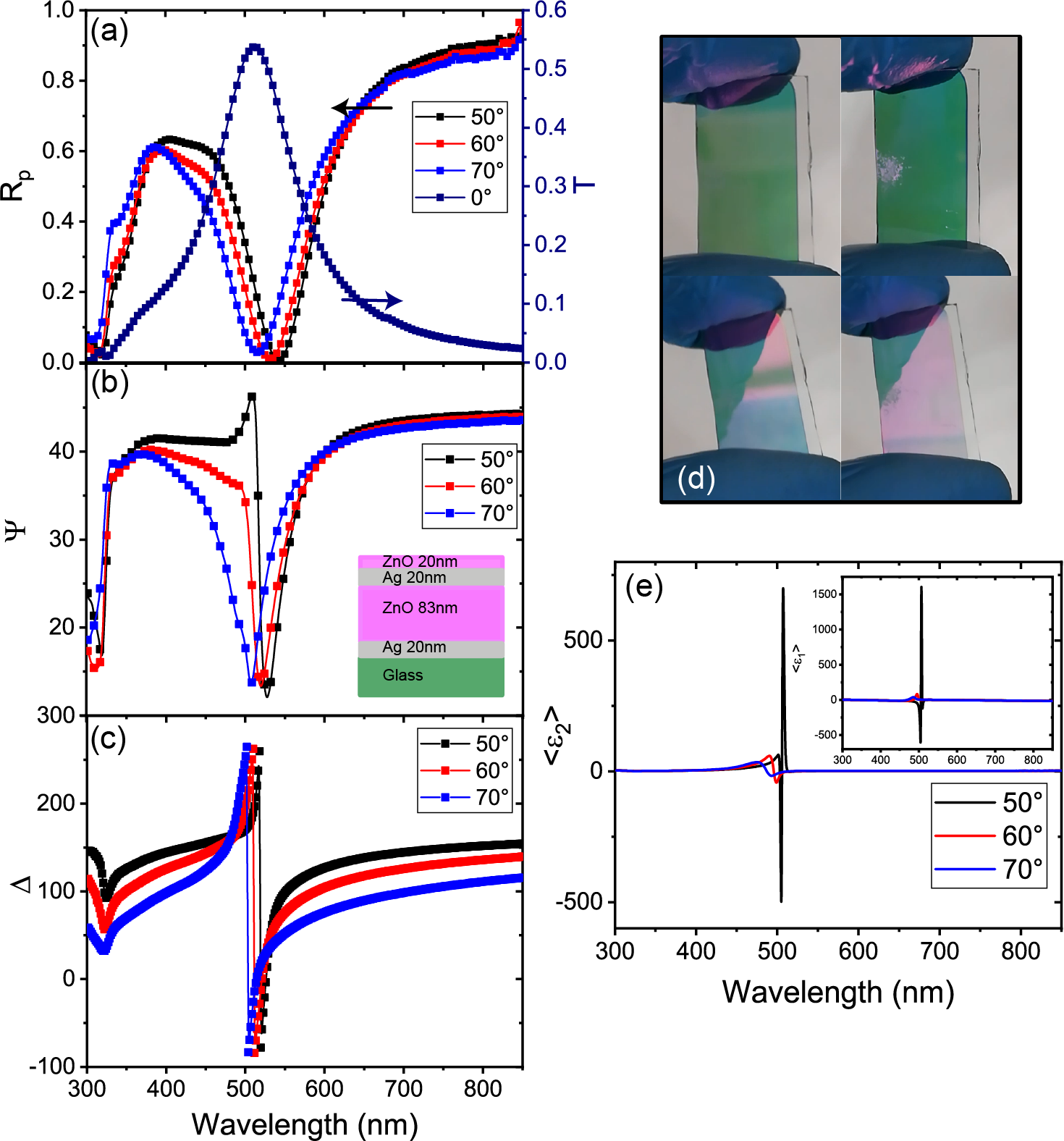}
\caption {a) Reflectances (p-pol), and transmittance at different incident angles for the MIMI composed of Ag/ZnO/Ag/ZnO with the ZnO layer of $t_{1}=83nm$ and $t_{2}=20nm$, respectively. b,c) $\Psi$ and $\Delta$ curves for the fabricated nano-cavity, the inset reports a schematic view of the MIMI with the related thickness of each layer. d) Pictures referred to Ag/ZnO/Ag/ZnO system taken at different incident/view angles. e) $<\varepsilon_2>$ at three different incident angles, while the inset shows the $<\varepsilon_1>$ curve.}
\label{4}
\end{center}
\end{figure}

Successively, in order to enhance the performance of the plasmonic MIMI nano-cavities,the ZnO has been used as dielectric instead of ITO. Moreover, ZnO in nano-science is largely used thanks to its excellent features in photovoltaics \cite{lee2008zno, manor2011electrical}, extreme non linearity application as second harmonic generation \cite{tritschler2003evidence, neumann2005second},  photocatalysis \cite{yayapao2013ultrasonic, jaramillo2004combinatorial}, just to name a few. 
The nano-cavity is composed of Ag/ZnO/Ag/ZnO and the dielectric layers have the following thickness $t_{1}=83nm$ and $t_{2}=20nm$, see the inset in Fig. \ref{4}. The proposed MIMI presents the same optical properties of the previous one: experimentally measured transmittance reaching the 55$\%$ at the resonant $\lambda=530nm$, and 0$\%$ reflectance at around ($\lambda=520nm$), see Fig. \ref{4}a .
As for the systems with ITO, the GSP and SPP arised inside the ZnO MIMI with the s-polarized reflectance (Rs) presenting a dip close to $0\%$ at the resonance wavelength ($\lambda=530nm$), see Fig. S6a. These features ($Rp$ and $Rs=0\%$) allow measuring a very sharp $\Psi$ and $\Delta$, as illustrated in Figs. \ref{4}b, c;  paving the way to exploit the enhanced Goos-H\"anchen shift for sensing applications.
This nano-cavity MIMI, due to the accentuate behavior of $\Psi$ and $\Delta$ at each incident angles, presents a remarkable angle color change passing from green color to some shades of purple. In comparison to the ITO MIMIs, the resonance of this system allows transmitting the green component of the light, while the other produce a purple reflection, as it is visible in the pictures of Fig. \ref{4}d. This ZnO system works better as filtering component at the resonant wavelength due to the less influence of the Ferrel-Berreman mode. The measured enhanced pseudo $<\varepsilon_2> = |500|$ at $\lambda=530nm$ for the incident angle $50^\circ$ correspond to main excited cavity modes (the high and low energy), see Fig. \ref{4}e. Moreover, at the other incident angles ($60^\circ$, $70^\circ$) gain resonant mode are still present confirmed by the measured sigmoidal shape of $\Delta$ parameter.  
The color change as a function of the incident/view angles is video recorded and available in the supplementary information. All the measured optical properties agree very well with the numerical simulations, which are reported in Figs. S6b,c. The electric field maps at and out of resonance wavelength for that ZnO MIMI has been numerically illustrated in Fig. S7.
The proposed systems allow realizing sub-wavelength plasmonic cavities without the use of complicate or nano-momentum matching structures. They can be exploited for future applications as sensors, especially considering the giant enhanced Goos-H\"anchen shift and its effects \cite{hashimoto1989optical, yin2006goos, sreekanth2019phase}, photovoltaics layers, plasmonic color filter for CMOS sensors \cite{yokogawa2012plasmonic}, plasmonic for multiple uses \cite{roberts2014subwavelength}, active nano-optics \cite{maier2006gain, lal2007nano, karalis2009plasmonic, martin2010resonance, ellenbogen2012chromatic}. They can also be exploited for the future physical security systems, for example it is possible to use the $\Psi$ or $\Delta$ optical response (or the pseudo dielectric constats due to the sharp curves) as a fingerprint for unclonable and anti-counterfeiting applications. \cite{cui2014encoding, zheng2016unclonable, smith2016plasmonic}.
  
\section{Conclusions}
In this experimental and numerical work, we have presented a comprehensive study of different optical metal/insulator nano-cavities. The systems show particular optical features such as zero reflectances, extraordinary transmittance, sigmoidal behavior of the parameters $\Psi$ and $\Delta$, and the giant enhanced real and imaginary part of the pseudo dielectric constants $<\varepsilon_1>$ and $<\varepsilon_2>$ at the resonant wavelengths. The MIMI configuration due to the double interface metal-insulator the dispersion curve exhibit high and low-energy modes which can be excited in free space without momentum matching. Hence, all the optical properties possesses by the proposed MIMI nano-cavities systems allow using them as sensors due to the giant enhanced Goos-H\"anchen shift. Furthermore, the color changing behavior could be exploited for filtering, physical security systems, and photonics.

\section{Materials and Methods}
\textbf{Samples fabrication} Fabrication began depositing silver on glass by using DC sputtering (power $30W$ and argon pressure of $4.5x10^{-2} mbar$ ), see the inset of Fig. \ref{1}a. Then, for the Ag/PVP/Ag nano-cavities, on the first Ag layer of $20nm$ the PVP 5$\%$ wt. in ethanol is deposited  by spin-coating (3000 rpm) resulting in a layer of $380nm$ measured by atomic force microscopy. Finally, the fabrication of the system is completed depositing silver ($20nm$) on top of PVP. A schematic view is reported in the inset of Fig. \ref{1}b. For the MIMIs composed of Ag/ITO/Ag/ITO the PVP is replaced with a thin layer of ITO deposited by DC sputtering (power $40W$ and argon pressure of $4.5x10^{-2} mbar$). The thickness of each layer of nano-cavity systems is reported in the inset of Fig. \ref{2}. The last MIMI is realized with ZnO instead of ITO. \\
\textbf{Characterization} The fabricated systems is characterized using a Wase-M2000 Woolam Ellipsometer from $300nm$ to $900nm$ with a resolution of 1.5nm in both p and s-polarization with incident angles: $50^\circ$, $60^\circ$, $70^\circ$. The experimental data is plotted with a skip point of 7. The measurements are fitted with a property software. In order to extracted optical constants of the used materials and systems, for each deposition, a reference sample is fabricated and analyzed (single layer of each material).\\
\textbf{Numerical simulations} The simulations are performed using Numerical Ellipsometer Analysis (NEA) implemented in commercial software COMSOL Multiphysics, see ref [$39$] for details. 
 
\begin{acknowledgement}
The authors thank the ``Area della Ricerca di Roma 2", Tor Vergata, for the access to the ICT Services (ARToV-CNR) for the use of the COMSOL Multiphysics Platform, Origin Lab and Matlab, and the Infrastructure ``BeyondNano" (PONa3-00362) of CNR-Nanotec for the access to research instruments.
\end{acknowledgement}
\bibliography{bibliograph}
\end{document}


\maketitle
\begin{abstract}
This work presents numerical and experimental results of plasmonic Metal-Insulator-Metal-Insulator nano-cavities with high gain effect. We use Silver as metal and polyvinylpyrrolidone or indium tin oxide or zinc oxide as insulator. The proposed systems exhibit extraordinary effects as colors changing in function of the incident/view angles and a giant enhancement of the experimental real and imaginary parts of the pseudo dielectric constant, a transmission of 50$\%$ and zero reflection for both polarizations at the resonant wavelengths. These results confirm the formation of surface plasmon polaritons and their hybridization in gap surface plasmons. The concurrence of these effects allow the study of the Goos-H\"anchen shift by exploiting the very narrow $\Delta$ and $\Psi$ parameters. Thank to their optical properties, the proposed nano-cavities could be applied in several fields such tunable color filters, active optics, photonics and sensors. 
\end{abstract}

\section{} 
\begin{figure}[H]
\begin{center}
\includegraphics[width=0.5\columnwidth]{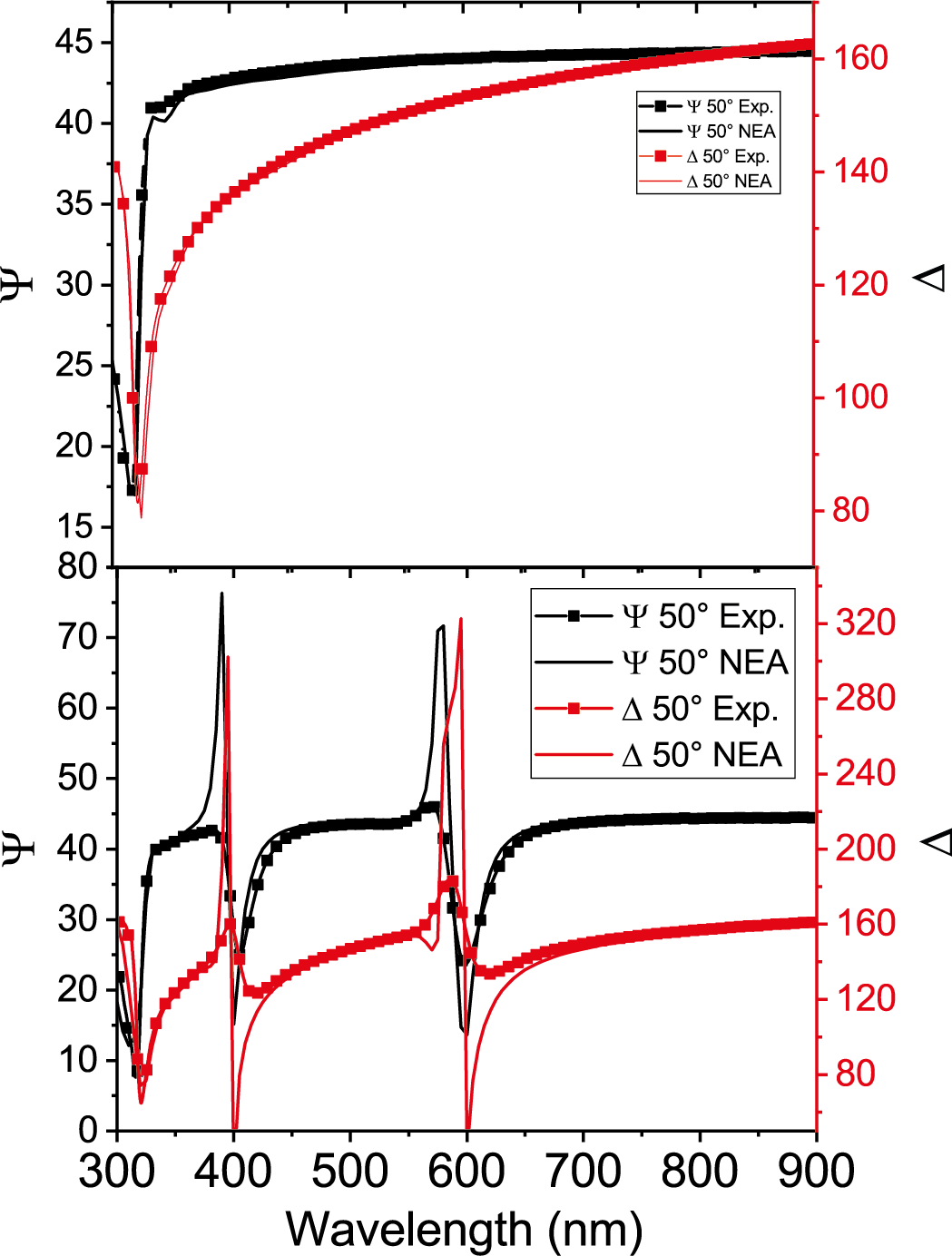}
  \caption{a) Experimental (Exp.) and numerical (NEA) $\Psi$ and $\Delta$ of the silver layer ($t=40nm$). b)  Experimental (Exp.) and numerical (NEA)  $\Psi$ and $\Delta$ of the nano cavity composed of PVP ($t=380nm$).}
  \label{1}
  \end{center}
\end{figure}

\begin{figure}[H]
\begin{center}
\includegraphics[width=0.5\columnwidth]{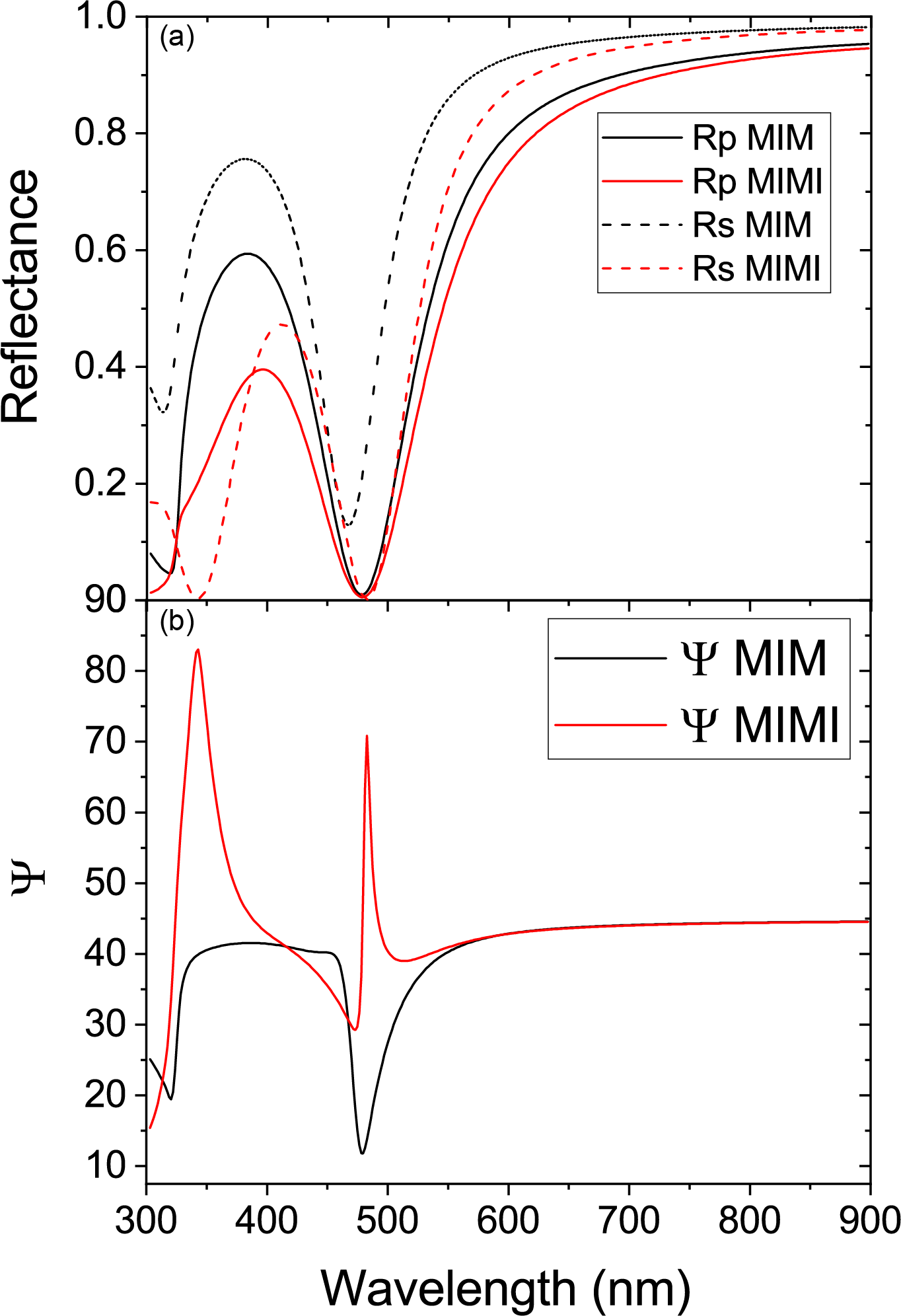}
  \caption{a,b) show the numerical reflectances, $\Psi$ and $\Delta$ for the MIM and MIMI. }
  \label{2_pre}
  \end{center}
\end{figure}

\begin{figure}[H]
\begin{center}
\includegraphics[width=0.5\columnwidth]{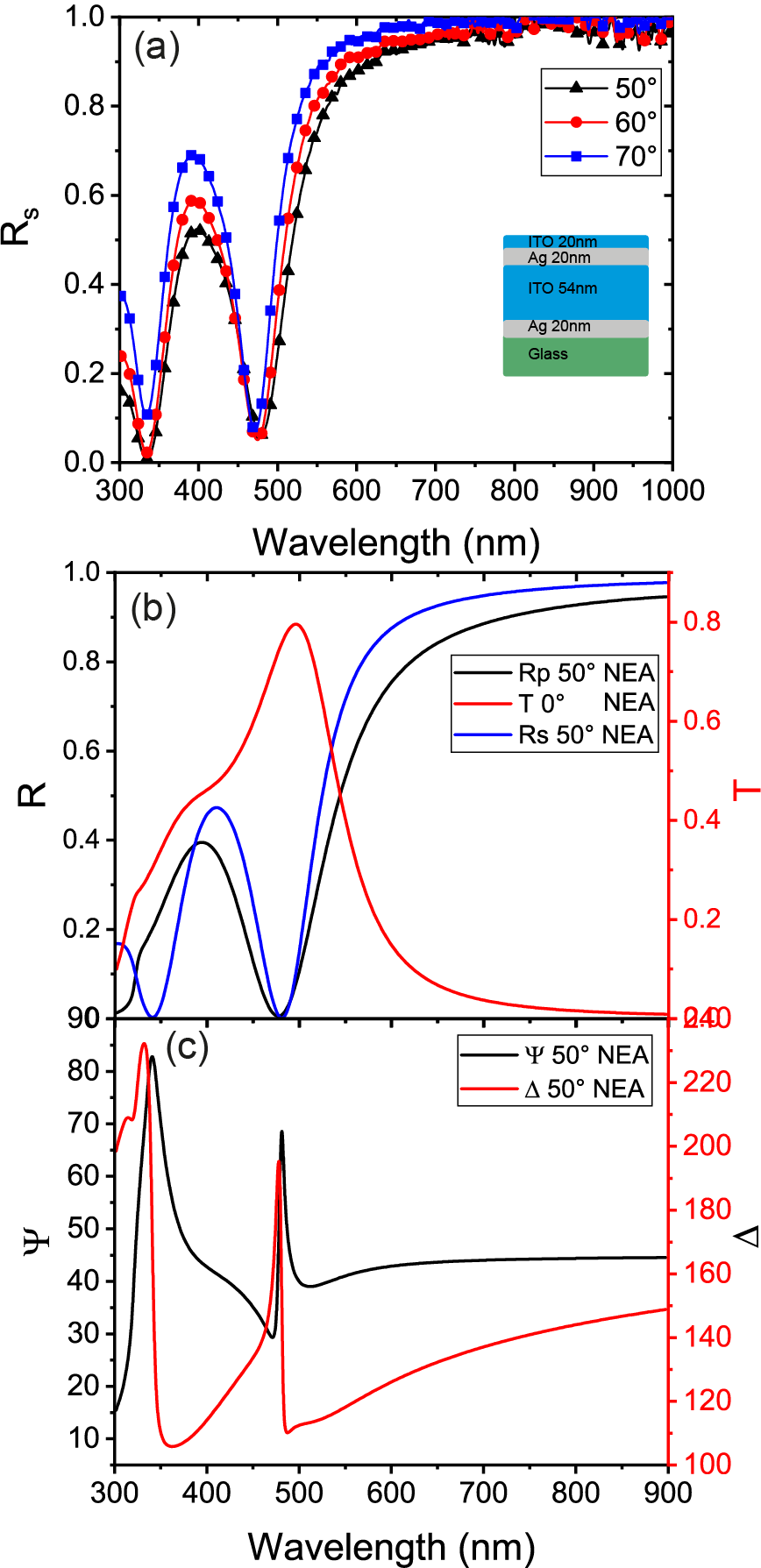}
  \caption{a) Reflectance (s-pol) at different incident angles for the system made by the ITO slab of $65nm$. b,c) Numerical results for reflectances (p and s-pol) and transmittance, $\Psi$ and $\Delta$.  }
  \label{2}
  \end{center}
\end{figure}

Fig\ref{1}a reports experimental and numerical ellipsometer parameters $\Psi$ and $\Delta$ for the silver layer ($t=40nm$),  while Fig\ref{1}b for nano-cavity composed of PVP as insulator.  
The Fig \ref{2_pre}a,b shows the numerical comparison btween the MIM and MIMI reflectances for the p and s-pol, $\Psi$ and $\Delta$ respectively. The presence of a s-polarized reflectance reaching 0$\%$ for the MIMI leads to a particular $\Psi$ curve. Indeed, it presents a minimum and maximum corresponding to the s-pol and p-pol, respectively.  
The Fig \ref{2} shows the experimental reflectance for the s-pol, and the numerical simulations for the reflectances (both polarizations), transmittance and the ellipsometer parameters for the system composed of an ITO layer of $65nm$. 

\begin{figure}[H]
\begin{center}
\includegraphics[width=0.5\columnwidth]{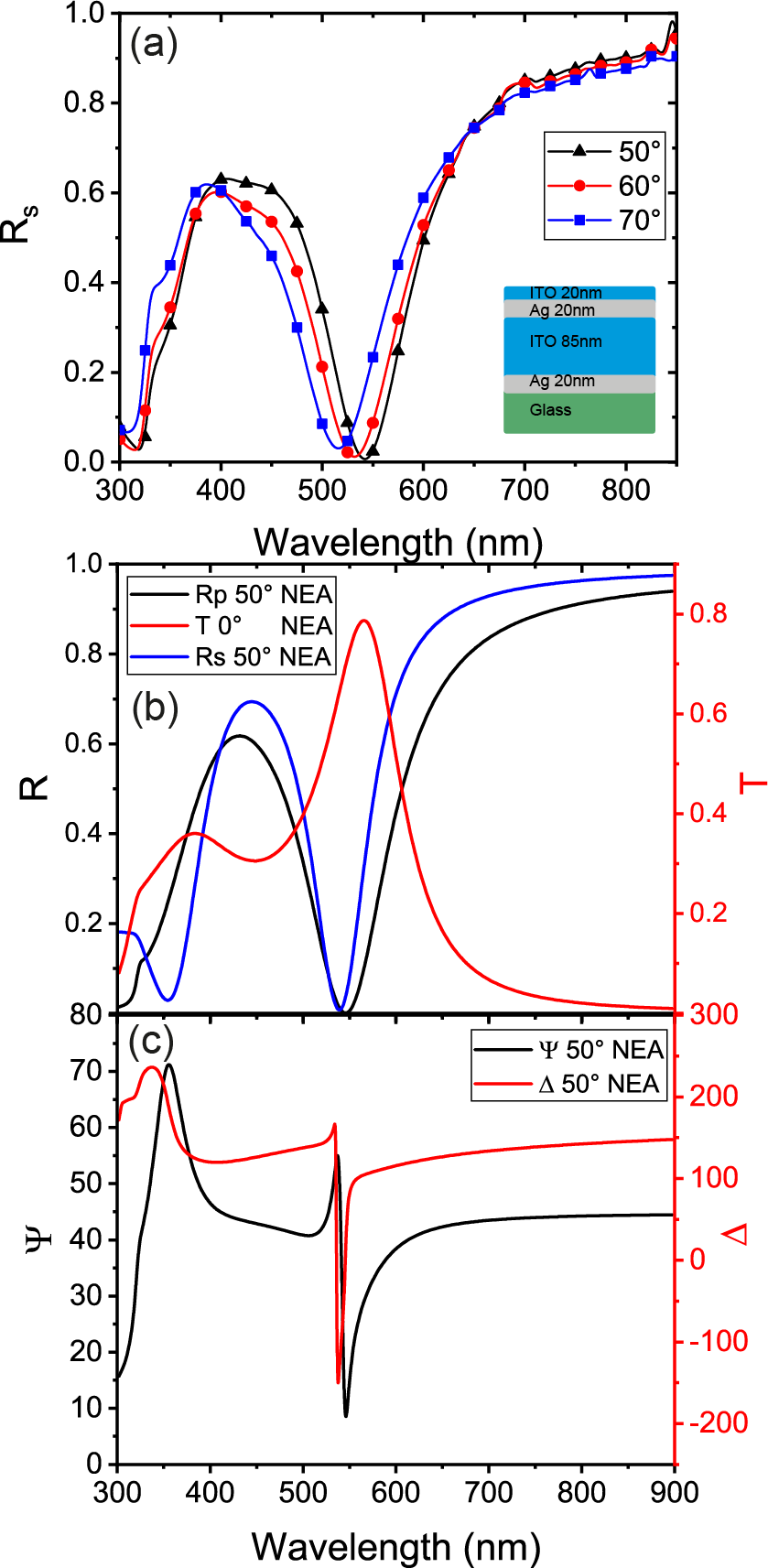}
  \caption{a) Reflectance (s-pol) at different incident angles for the system made by the ITO slab of $85nm$. b,c) Numerical results for reflectances (p and s-pol) and transmittance, $\Psi$ and $\Delta$. }
  \label{3}
  \end{center}
\end{figure}

\begin{figure}[H]
\begin{center}
\includegraphics[width=0.6\columnwidth]{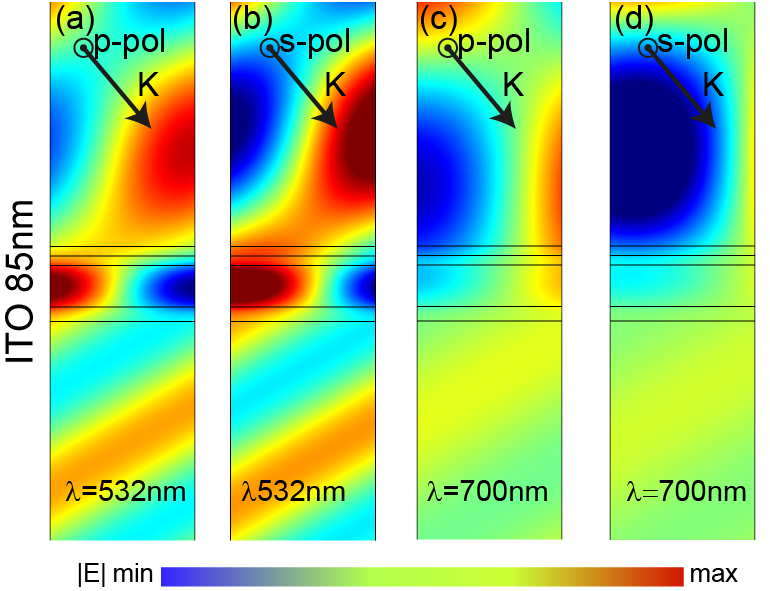}
  \caption{Electric field maps at and out of the resonance for the system composed of an ITO layer of $85nm$. a, b) Electric filed maps for p and s-polarization at $\lambda=532nm$. c, d) The out of resonance for both polarization at $\lambda=700nm$. }
  \label{3bis}
  \end{center}
\end{figure}

\begin{figure}[H]
\begin{center}
\includegraphics[width=0.5\columnwidth]{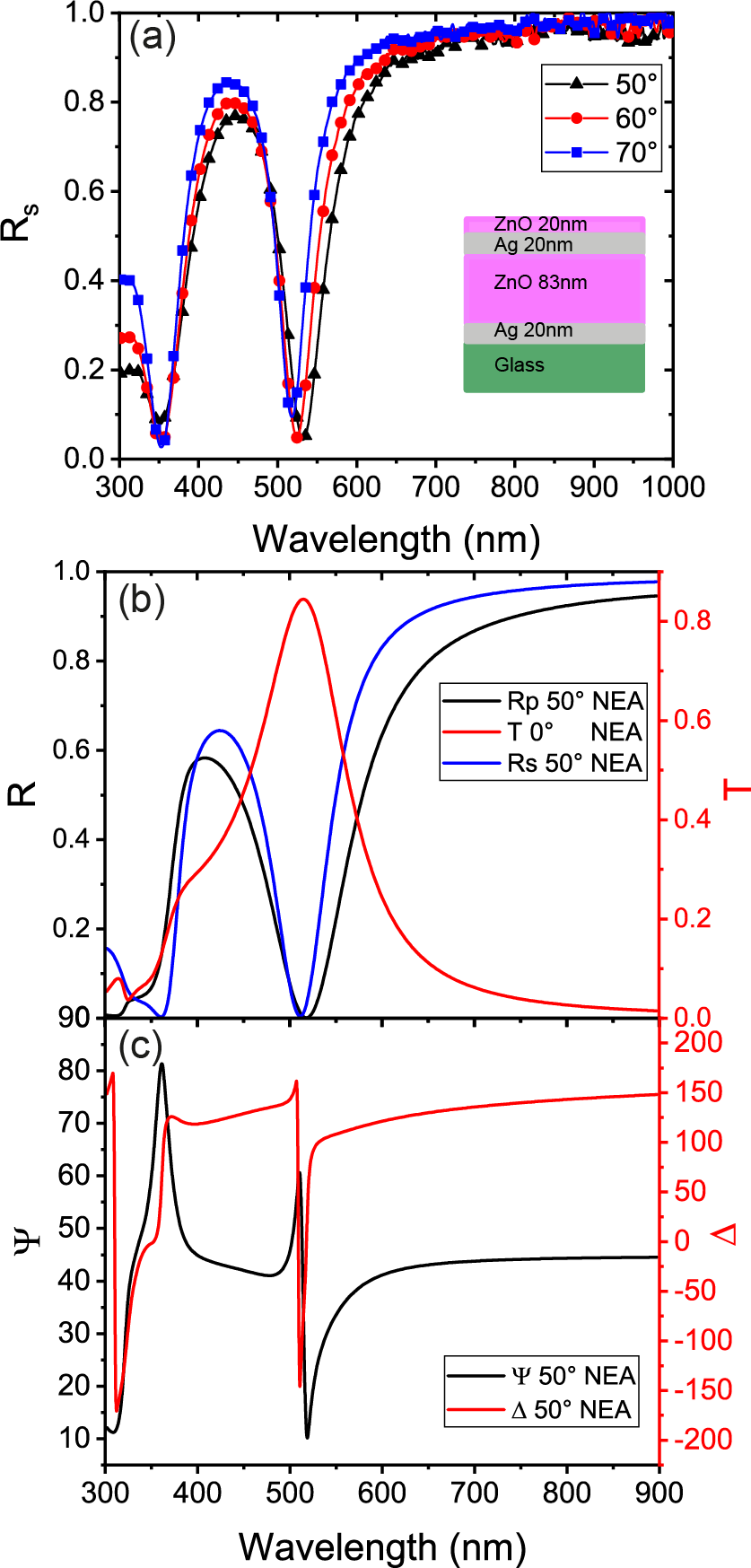}
\caption{a) Reflectance (s-pol) at different incident angles for the system composed of the ZnO slab of $83nm$. b,c) Numerical results for reflectances (p and s-pol) and transmittance, $\Psi$ and $\Delta$. }
\label{4}
\end{center}
\end{figure}

\begin{figure}[H]
\begin{center}
\includegraphics[width=0.6\columnwidth]{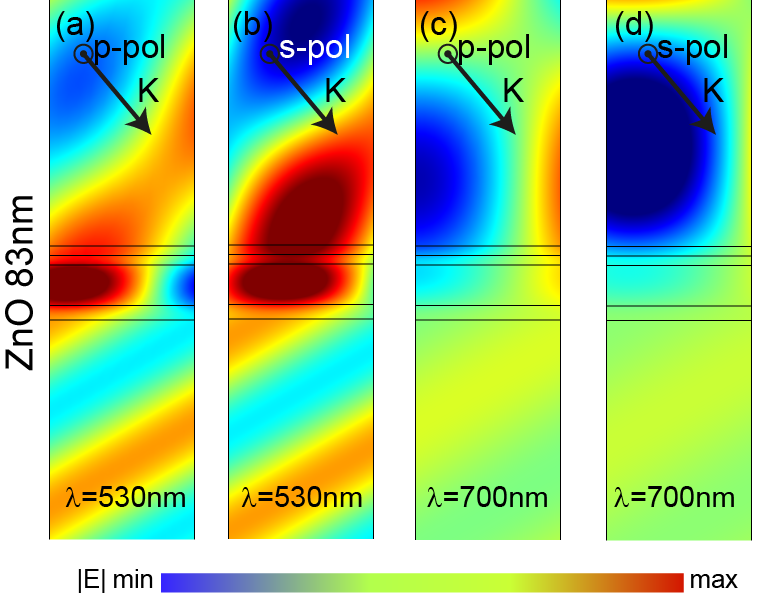}
\caption{Electric field maps at and out of the resonance for the system composed of a ZnO layer of $83nm$. a , b) The electric filed maps for p and s-polarization at $\lambda=530nm$. c, d) The out of resonance for both polarization at $\lambda=700nm$. }
\label{4bis}
\end{center}
\end{figure}

Fig \ref{3} shows the experimental reflectance for the s-pol, and the numerical simulations for the reflectances (both polarizations), transmittance and the ellipsometer parameters for the system composed of an ITO layer of $85nm$. 
Fig \ref{3bis} shows the electric field maps from an incident electromagnetic wave at 50 degree that impinges on the system composed of a ITO layer of $85nm$. We analyzed the system in two main conditions at and out of the resonance.
Fig \ref{4} shows the experimental reflectance for the s-pol, and the numerical simulation about the reflectance (both polarizations) transmittance and the ellipsometer parameters for a system composed of a ZnO slab of $83nm$. 
 Fig \ref{4bis} shows the electric field maps from an incident electromagnetic wave at 50 degree that impinges on the system composed of a ZnO layer of $83nm$. We analyzed the system in two main conditions at and out of the resonance.

\section{Acknowledgment}
The authors thank the ``Area della Ricerca di Roma 2", Tor Vergata, for the access to the ICT Services (ARToV-CNR) for the use of the COMSOL Multiphysics Platform, Origin Lab and Matlab, and the Infrastructure ``BeyondNano" (PONa3-00362) of CNR-Nanotec for the access to research instruments.

\bibliography{bibliograph}